\documentclass[a4paper,11pt]{article}
\usepackage{amssymb,amsmath,amsfonts,placeins,pbox,multirow,mathtools}
\usepackage{graphicx,rotate,color,slashed,cite,epstopdf,verbatim,url,multirow,booktabs}
\usepackage{epstopdf}
\usepackage{bm}
\usepackage{graphicx,caption}
\usepackage{array}
\usepackage[T1]{fontenc}
\renewcommand{\arraystretch}{1.5} 
\usepackage[a4paper, left=2cm, right=2cm, top=2.5cm, bottom=2.5cm]{geometry}
\usepackage[colorlinks=true,
linkcolor=magenta,
urlcolor=blue,
citecolor=blue]{hyperref}

\numberwithin{equation}{section}

\newcommand{\mt}{m_t}
\newcommand{\mh}{m_h}
\newcommand{\ms}{m_*}
\newcommand{\dd}{\mathrm d}

\newcommand{\dif}{\mathrm d}
\newcommand{\ii}{\mathrm i}

\newcommand{\Li}{\operatorname{Li}_2}

\newcommand{\cM}{\mathcal M}
\newcommand{\ttbar}{t\bar t}

\newcommand{\Gzero}{\Gamma_0}

\newcommand{\repl}[1]{{\color{red}#1}}

\def\ssbh#1//#2//{\ensuremath{\xrightarrow [\substack{#2}]
		{\parbox{3cm}{\hfil $\scriptstyle \langle #1 \rangle$ \hfil}}}}

\let\CapTion=\caption
\def\caption#1{\CapTion{\em #1}}

\usepackage[capitalise]{cleveref}
\crefname{section}{Sec.}{Secs.}
\crefname{table}{Tab.}{Tabs.}
\crefname{figure}{Fig.}{Figs.}
\crefname{equation}{Eq.}{Eqs.}
\crefname{appendix}{Appendix}{Appendix}

\title{\bf  Imprints of Non-local Top-Higgs Interaction on \\Loop Induced Higgs Processes}

\author{\bf Avik Banerjee$^{a}$\thanks{avik92kol@gmail.com}\,, Rohan Pramanick$^{b,c,d}$\thanks{rohanpramanick25@gmail.com}\,, Tirtha Sankar Ray$^{e}$\thanks{tirthasankar.ray@gmail.com}
	\\ \medskip 
	{}$^{a}$\small\em Department of Theoretical Physics, Tata Institute of Fundamental Research,\\[-13pt] \small\em Homi Bhabha Road, Mumbai 400005, India\\ 
	{}$^{b}$\small\em School of Physical Sciences, Indian Association for the Cultivation of Science, \\[-10pt]\small\em Kolkata-700032, India\\ 
	{}$^{c}$\small\em State Key Laboratory of Dark Matter Physics, Tsung-Dao Lee Institute \& School of \\[-10pt]\small\em Physics and Astronomy, Shanghai Jiao Tong University, Shanghai 200240, China\\
	{}$^{d}$\small\em Key Laboratory for Particle Astrophysics and Cosmology (MOE) \& Shanghai Key Laboratory \\[-10pt]\small\em for Particle Physics and Cosmology, Shanghai Jiao Tong University, Shanghai 200240, China\\
	{}$^{e}$\small\em Department of Physics, Indian Institute of Technology Kharagpur, Kharagpur 721302, India
}
\date{}

\usepackage{subcaption, soul}
\usepackage{tikz}
\usepackage[compat=1.1.0]{tikz-feynman}
\usetikzlibrary{patterns, patterns.meta}
\usetikzlibrary{arrows, arrows.meta}
\usetikzlibrary{shapes, shapes.multipart}
\usetikzlibrary{decorations.pathreplacing, decorations.pathmorphing, decorations.markings}
\usetikzlibrary{trees, matrix, calc, through, positioning}
\tikzset{
	blob_1/.style = {
		draw = black, thick, circle, minimum size = 6mm, inner sep = 1pt, outer sep = 1pt, fill = white,
		postaction = {
			pattern = {Lines[angle = 45, distance = 2pt, line width = 0.5pt]}
		}
	},
	blob_2/.style = {
		draw = black, thick, circle, minimum size = 6mm, inner sep = 1pt, outer sep = 1pt, fill = white,
		postaction = {
			pattern = {Lines[angle = 135, distance = 2pt, line width = 0.5pt]}
		}
	},
	blob_3/.style = {
		draw = black, thick, circle, minimum size = 6mm, inner sep = 1pt, outer sep = 1pt, fill = white,
		postaction = {
			pattern = crosshatch
		}
	},
}

\def\feyndiagFFtthOne{
	\scalebox{0.75}{
		\begin{tikzpicture}[baseline=(current bounding box.center)]
			\begin{feynman}
				\vertex (a) {\large $h (q)$};
				\vertex [right = 2cm of a] (b);
				\vertex [above right = 1.5cm and 1.5cm of b] (c);
				\vertex [below right = 1.5cm and 1.5cm of b] (d);
				\vertex [right = 2cm of c] (e) {\large $g(l_1)$};
				\vertex [right = 2cm of d] (f) {\large $g(l_2)$};
				\diagram{
					(a) -- [scalar, very thick] (b) -- [anti fermion, very thick, edge label = { $k + l_1$}] (c) -- [gluon, very thick] (e);
					(b) -- [fermion, very thick, edge label' = {$k - l_2$}] (d) -- [gluon, very thick] (f);
					(c) -- [anti fermion, very thick, edge label = {\large $k$} ] (d) ;
				};
				\node   [right = 2cm of a, blob_1] (b1);
				
				\vertex [above left = 0.35cm and 0.65cm of e] (e1);
				\vertex [above right= 0.35cm and 0.45cm of c] (c1);
				\draw[->, >=stealth, line width=1.5pt] (e1) -- (c1);
				\vertex [below left = 0.35cm and 0.65cm of f] (f1);
				\vertex [below right= 0.35cm and 0.45cm of d] (d1);
				\draw[->, >=stealth, line width=1.5pt] (f1) -- (d1);
				\vertex [below left = 0.35cm and 0.45cm of b] (b1);
				\vertex [below right= 0.35cm and 0.45cm of a] (a1);
				\draw[->, >=stealth, line width=1.5pt] (b1) -- (a1);
			\end{feynman}
		\end{tikzpicture}
	}
}

\def\feyndiagFFtthTwo{
	\scalebox{0.75}{
		\begin{tikzpicture}[baseline=(current bounding box.center)]
			\begin{feynman}
				\vertex (a) {\large $h (q)$};
				\vertex [right = 2cm of a] (b);
				\vertex [above right = 1.5cm and 1.5cm of b] (c);
				\vertex [below right = 1.5cm and 1.5cm of b] (d);
				\vertex [right = 2cm of c] (e) {\large $g(l_2)$};
				\vertex [right = 2cm of d] (f) {\large $g(l_1)$};
				\diagram{
					(a) -- [scalar, very thick] (b) -- [anti fermion, very thick, edge label = { $k + l_2$}] (c) -- [gluon, very thick] (e);
					(b) -- [fermion, very thick, edge label' = {$k - l_1$}] (d) -- [gluon, very thick] (f);
					(c) -- [anti fermion, very thick, edge label = {\large $k$} ] (d) ;
				};
				\node   [right = 2cm of a, blob_1] (b1);
				
				\vertex [above left = 0.35cm and 0.65cm of e] (e1);
				\vertex [above right= 0.35cm and 0.45cm of c] (c1);
				\draw[->, >=stealth, line width=1.5pt] (e1) -- (c1);
				\vertex [below left = 0.35cm and 0.65cm of f] (f1);
				\vertex [below right= 0.35cm and 0.45cm of d] (d1);
				\draw[->, >=stealth, line width=1.5pt] (f1) -- (d1);
				\vertex [below left = 0.35cm and 0.45cm of b] (b1);
				\vertex [below right= 0.35cm and 0.45cm of a] (a1);
				\draw[->, >=stealth, line width=1.5pt] (b1) -- (a1);
			\end{feynman}
		\end{tikzpicture}
	}
}

\def\feyndiagFFtthgOne{
	\scalebox{0.75}{
		\begin{tikzpicture}[baseline=(current bounding box.center)]
			\begin{feynman}
				\vertex (a) {\large $h (q)$};
				\vertex [right = 2.75cm of a] (b);
				\vertex [below = 3.0cm of b] (c);
				\vertex [right = 2.75cm of b] (d) {\large $g(l_1)$};
				\vertex [right = 2.75cm of c] (e) {\large $g(l_2)$};
				\diagram{
					(a) -- [scalar, very thick] (b) -- [gluon, very thick] (d);
					(c) -- [gluon, very thick] (e);
					(b) -- [fermion, very thick, half left, looseness = 1, edge label = {$k + l_2$}] (c);
					(c) -- [fermion, very thick, half left, looseness = 1, edge label = {$k$}] (b)
				};
				\node   [right = 2.75cm of a, blob_2] (b1);
				
				\vertex [above left = 0.35cm and 0.65cm of d] (d1);
				\vertex [above right= 0.35cm and 0.45cm of b] (b1);
				\draw[->, >=stealth, line width=1.5pt] (d1) -- (b1);
				\vertex [below left = 0.35cm and 0.65cm of e] (e1);
				\vertex [below right= 0.35cm and 0.45cm of c] (c1);
				\draw[->, >=stealth, line width=1.5pt] (e1) -- (c1);
				\vertex [below left = 0.35cm and 0.85cm of b] (b1);
				\vertex [below right= 0.35cm and 0.45cm of a] (a1);
				\draw[->, >=stealth, line width=1.5pt] (b1) -- (a1);
			\end{feynman}
		\end{tikzpicture}
	}
}

\def\feyndiagFFtthgTwo{
	\scalebox{0.75}{
		\begin{tikzpicture}[baseline=(current bounding box.center)]
			\begin{feynman}
				\vertex (a) {\large $h (q)$};
				\vertex [right = 2.75cm of a] (b);
				\vertex [below = 3.0cm of b] (c);
				\vertex [right = 2.75cm of b] (d) {\large $g(l_2)$};
				\vertex [right = 2.75cm of c] (e) {\large $g(l_1)$};
				\diagram{
					(a) -- [scalar, very thick] (b) -- [gluon, very thick] (d);
					(c) -- [gluon, very thick] (e);
					(b) -- [fermion, very thick, half left, looseness = 1, edge label = {$k + l_1$}] (c);
					(c) -- [fermion, very thick, half left, looseness = 1, edge label = {$k$}] (b)
				};
				\node   [right = 2.75cm of a, blob_2] (b1);
				
				\vertex [above left = 0.35cm and 0.65cm of d] (d1);
				\vertex [above right= 0.35cm and 0.45cm of b] (b1);
				\draw[->, >=stealth, line width=1.5pt] (d1) -- (b1);
				\vertex [below left = 0.35cm and 0.65cm of e] (e1);
				\vertex [below right= 0.35cm and 0.45cm of c] (c1);
				\draw[->, >=stealth, line width=1.5pt] (e1) -- (c1);
				\vertex [below left = 0.35cm and 0.85cm of b] (b1);
				\vertex [below right= 0.35cm and 0.45cm of a] (a1);
				\draw[->, >=stealth, line width=1.5pt] (b1) -- (a1);
			\end{feynman}
		\end{tikzpicture}
	}
}

\def\feyndiagFFtthgg{
	\scalebox{0.75}{
		\begin{tikzpicture}[baseline=(current bounding box.center)]
			\begin{feynman}
				\vertex (a) {\large $h (q)$};
				\vertex [right = 2.75cm of a] (b);
				\vertex [above right = 1.5cm and 2.75cm of b] (c) {\large $g(l_1)$};
				\vertex [below right = 1.5cm and 2.75cm of b] (d) {\large $g(l_2)$};
				\vertex [above left = 1.5cm and 0.5cm of b] (e);
				\diagram{
					(a) -- [scalar, very thick] (b) -- [gluon, very thick] (c);
					(b) -- [gluon, very thick] (d);
					(b) -- [fermion, half left, very thick, looseness = 1.25, edge label' = {$k$}] (e);
					(e) -- [fermion, half left, very thick, looseness = 1.25] (b);
				};
				\node   [right = 2.75cm of a, blob_3] (b1);
				
				\vertex [below left = 0.75cm and 0.65cm of c] (c1);
				\vertex [above right= 0.35cm and 1.25cm of b] (b1);
				\draw[->, >=stealth, line width=1.5pt] (c1) -- (b1);
				\vertex [above left = 0.75cm and 0.65cm of d] (d1);
				\vertex [below right= 0.35cm and 1.25cm of b] (b2);
				\draw[->, >=stealth, line width=1.5pt] (d1) -- (b2);
				\vertex [below left = 0.35cm and 0.85cm of b] (b1);
				\vertex [below right= 0.35cm and 0.45cm of a] (a1);
				\draw[->, >=stealth, line width=1.5pt] (b1) -- (a1);
			\end{feynman}
		\end{tikzpicture}
	}
}

\begin{document}
	
	\maketitle

	% % % % % % % % % % % % % % % % Abstract % % % % % % % % % % % % % % % % % 
	
	\begin{abstract}
		
		A composite Higgs arising from an underlying strong sector generally has non-local interactions with the elementary sector, formulated in terms of momentum-dependent form factors.  We construct the minimal Ball-Chiu-type vertices consistent with gauge-invariant non-local $t\bar{t}h$ couplings containing Wilson lines, where recursive Ward-Takahashi identities connect the Yukawa interaction to higher-order contact interactions. For a symmetric monopole form factor motivated by lowest-resonance dominance, we compute the top-loop contributions to $gg \to h$ and $h \to \gamma\gamma$ amplitudes in closed form, and demonstrate a logarithmically enhanced deviation from the Standard Model. Current measurements at the LHC exclude the characteristic new physics scale below 3.6 TeV at $2\sigma$, while future colliders could extend the sensitivity to approximately 8 TeV.

	\end{abstract}
	
	% % % % % % % % % % % % % % % % % % % % % % % % % % % % % % % % % % % % % %
	
	\bigskip
	
	% % % % % % % % % % % % % % % Main Document% % % % % % % % % % % % % % % % 
	\newpage
	{
		\hypersetup{linkcolor=black}
		\tableofcontents
	}
	
	\bigskip
	\hrule
	\bigskip

	% % % % % % % % % % % % % % % % % % % % % % % % % % % % % % % % % % % % % %
	\section{Introduction}
	\label{intro}

	The discovery of the $\sim 125~\text{GeV}$ Higgs boson at the Large Hadron Collider (LHC)~\cite{ATLAS:2012yve, CMS:2012qbp} has transformed the study of the electroweak scalar sector  into a field of  precision tests for the Standard Model (SM) and beyond. Over the course of LHC Runs~2 and~3, the ATLAS and CMS collaborations have measured the  Higgs signal strengths in the loop driven gluon fusion and di-photon channels to an accuracy of ten percent or better~\cite{ATLAS:2022vkf, CMS:2022dwd}. Looking ahead, the High-Luminosity LHC (HL-LHC)~\cite{Cepeda:2019klc} and proposed next-generation lepton and hadron colliders, such as the ILC~\cite{Bambade:2019fyw}, CEPC~\cite{CEPCStudyGroup:2018ghi}, and FCC~\cite{FCC:2018byv, FCC:2018evy}, are projected to push these measurements into the sub-percent regime~\cite{deBlas:2019rxi}. Despite the impressive agreement of current data with the SM predictions, coupling measurements near the electroweak scale do not by themselves settle whether the observed scalar is truly an elementary pointlike particle or the low-energy manifestation of a composite scalar arising from an underlying strong dynamics.
	
	The Composite Higgs paradigm as a resolution of the gauge hierarchy problem has received extensive attention in the literature~\cite{Kaplan:1983fs, Kaplan:1983sm, Kaplan:1991dc, Agashe:2004rs, Contino:2010rs, Bellazzini:2014yua, Panico:2015jxa, Barnard:2013zea, Ferretti:2013kya}. In this framework, a light Higgs boson emerges naturally as a pseudo-Nambu-Goldstone boson (pNGB) of a spontaneously broken global symmetry within a strongly coupled sector confining at a scale $m_* \sim \mathcal{O}(\text{TeV})$. Because the Higgs is a bound state of the underlying strong dynamics, its interactions with the rest of the SM carry imprints of the substructure and resonance spectrum of the composite sector \cite{Murayama:2014yja,Banerjee:2017qod,Goncalves:2018pkt,Goncalves:2020vyn,MammenAbraham:2021ssc, Banerjee:2105.01093, Bittar:2022pmq,Banerjee:2024puz}. In the present study, we focus on how the compositeness of the Higgs can be probed through its non-local coupling with the elementary top sector. Because the composite Higgs state possesses a finite spatial extent governed by the confinement scale $m_*$, its interaction with the elementary  states such as the top quarks cannot be pointlike at arbitrarily short distances. Integrating out the heavy strong sector resonances generates a non-local $htt$ interaction, replacing the pointlike SM Yukawa coupling with a momentum-dependent vertex function $\Gamma_0(p',p, m_*)$ of the incoming ($p$) and outgoing ($p'$) top-quark momenta~\cite{Agashe:2004rs, Montull:2013mla}. Comparable momentum-dependent top-Higgs vertex functions also arise when the Higgs sector resides near a non-trivial quantum critical point~\cite{Bellazzini:2015cgj}, in approximately scale-invariant or unparticle-like hidden sectors characterized by a continuum spectral density~\cite{Georgi:2007ek, Cacciapaglia:2007jq}, and in explicit ultraviolet (UV) completions where radiative exchanges of heavy states induce calculable Yukawa form factors~\cite{Bittar:2022pmq}.%~\cite{Bittar:2022wgb}.
	
	Deviations of the top--Higgs interaction from its SM form are commonly
	parameterized by a momentum independent coupling modifier $\kappa_t$,
	or described through a finite set of local higher dimensional operators
	within the Standard Model Effective Field Theory (SMEFT)~\cite{Giudice:2007fh}
	or Higgs Effective Field Theory
	(HEFT)~\cite{Buchalla:2013rka,Brivio:2016fzo,Brivio:2017vri}.
	While the former captures an overall rescaling of the Yukawa coupling,
	the latter allows a systematic derivative expansion of momentum-dependent
	interactions. Such a local description is reliable at energies well
	below the compositeness scale $m_*$, provided the effective operators
	are consistently matched and renormalized. Retaining the complete non-local form factor provides
	a handle to consistently capture processes where the momentum scan virtualities beyond the scale of the underlying strong dynamics. Loop-induced
	processes are particularly well suited to capture the full momentum dependence of the vertex function since the internal virtualities span over a wide range of scales. 
	
	A naive insertion of
	such a form factor into the three-point $\bar t t h$ vertex, however,
	generally violates the Ward--Takahashi identities (WTIs) in processes
	such as $gg\to h$. This follows from the non-local nature of the
	interaction, which involves top-quark fields evaluated at separated
	spacetime points. Restoring gauge invariance requires additional
	contact interactions, including $\bar t t hg$ and $\bar t t hgg$ vertices,
	related to the momentum-dependent Yukawa coupling through recursive
	WTIs~\cite{Haberzettl:1997jg}. These interactions can be systematically
	generated by connecting the separated top-quark fields with a
	path-ordered QCD Wilson line~\cite{Mandelstam:1962mi,Terning:1991yt}.
	Expanding the Wilson line in powers of $g_s$ yields the multi-gluon
	contact vertices necessary to maintain gauge invariance at successive
	orders. Analogously, electromagnetic gauge invariance requires an electromagnetic Wilson line,
	which generates the corresponding photon contact interactions
	satisfying the QED Ward--Takahashi identities.
	%\cite{Ohta:1989ji, Haberzettl:1997jg, Davidson:2001rk}
	
	The momentum dependent $\bar t t h$ form factor $\Gamma_0(p',p)$,
	together with the associated one- and two-gluon contact vertices
	$\Gamma_1^\mu$ and $\Gamma_2^{\mu\nu}$, forms a consistent
	gauge-invariant set. Following the Ball-Chiu
	prescription~\cite{Ball:1980ay}, the recursive WTIs admit a minimal
	longitudinal solution, while transverse contributions remain
	unconstrained by gauge invariance alone and depend on the chosen
	ultraviolet gauge completion. The loop-induced processes $gg\to h$~\cite{Georgi:1977gs}
	and $h\to\gamma\gamma$~\cite{Ellis:1975ap,Shifman:1979eb} provide
	complementary probes of the momentum dependence of the top--Higgs
	interaction. Since both processes arise at one loop in the SM,
	modifications of the top-loop amplitude enter at the same perturbative
	order as the SM contribution. In particular, $gg\to h$ probes the
	interplay between the conventional triangle diagrams and the
	$\bar t t hg$ and $\bar t t hgg$ contact interactions. A suppression
	of the top-loop contribution reduces the gluon-fusion production
	rate ($\mu_{gg}\equiv \sigma_{gg}/\sigma^{\rm SM}_{gg}<1$), while simultaneously weakening its destructive
	interference with the dominant $W$-boson loop in $h\to\gamma\gamma$,
	thereby enhancing the diphoton decay width
	($\mu_{\gamma\gamma}\equiv BR_{\gamma\gamma}/BR^{\rm SM}_{\gamma\gamma}>1$). 
	
	In this work, we develop a gauge-invariant formalism for non-local top--Higgs interactions and investigate their effects on these two processes. Adopting a symmetric monopole form factor motivated by lowest resonance dominance~\cite{Contino:2010rs,Banerjee:2017qod,Banerjee:2105.01093},
	we evaluate the complete top-loop contributions analytically in
	terms of standard Passarino--Veltman functions. Although individual
	triangle and contact diagrams contain ultraviolet divergences,
	these cancel in the complete gauge-invariant amplitude. We find
	a logarithmically enhanced deviation from the SM, scaling as
	$(m_t^2/m_*^2)\ln(m_*^2/m_t^2)$, reflecting the imprints of the momentum dependent vertex. 
	%For the gauge completion considered here, the transverse contribution associated with the two-gluon contact vertex scales as $m_t^2/m_*^2$ without logarithmic enhancement and is therefore parametrically subleading. 
	Confronting our predictions with current LHC measurements, we obtain a lower bound of approximately $3.6~\mathrm{TeV}$ on $m_*$ at the $2\sigma$ level. Projected sensitivities at the HL-LHC, ILC, CEPC, and FCC could extend this reach to the $7$--$8~\mathrm{TeV}$ range.
	
	The remainder of this paper is organized as follows. In Section~\ref{sec:form-factor}, we demonstrate the violation of gauge invariance under a naive form-factor insertion in $gg \to h$ amplitude and derive the recursive WTIs for the higher-order contact vertices. In Section~\ref{non-local}, we construct the minimal Ball-Chiu-type interaction and the non-local Wilson-line completion, analyzing the longitudinal and transverse components of the vertices and the conditions for contour independence. In Section~\ref{sec:illustrativeFF}, we classify representative form-factor parameterizations and detail the contact vertices for our symmetric monopole benchmark. In Section~\ref{pheno}, we present the closed-form one-loop amplitudes for $gg \to h$ and $h \to \gamma\gamma$, compare the non-local form-factor predictions against a constant $\kappa_t$ rescaling, and obtain the current and projected collider bounds on $m_*$. We summarize our conclusions in Section~\ref{sec:conclusion}.

	% % % % % % % % % % % % % % % % % % % % % % % % % % % % % % % % % % % % % %
	%==============================================================================
	\section{Form factor and gauge invariance}
	\label{sec:form-factor}
	%==============================================================================

\begin{figure}[t]
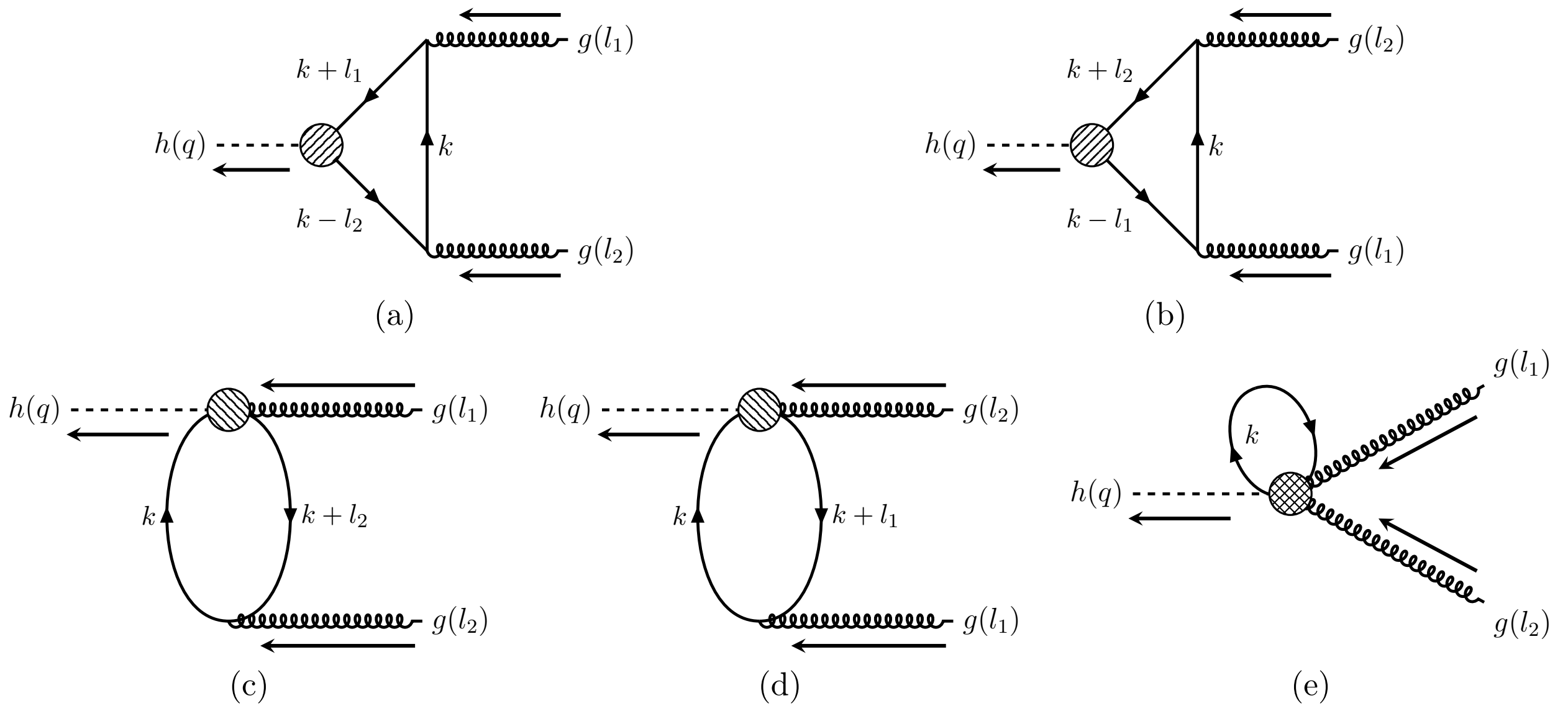

		\centering
		\begin{subfigure}{0.475\linewidth}
			\centering
			\feyndiagFFtthOne
			\caption{}
			\label{label_fig_tth_one}
		\end{subfigure}
		\begin{subfigure}{0.475\linewidth}
			\centering
			\feyndiagFFtthTwo
			\caption{}
			\label{label_fig_tth_two}
		\end{subfigure}\\
		\centering
		\begin{subfigure}{0.325\linewidth}
			\centering
			\feyndiagFFtthgOne
			\caption{}
			\label{label_fig_tthg_one}
		\end{subfigure}
		\begin{subfigure}{0.325\linewidth}
			\centering
			\feyndiagFFtthgTwo
			\caption{}
			\label{label_fig_tthg_two}
		\end{subfigure}
		\begin{subfigure}{0.325\linewidth}
			\centering
			\feyndiagFFtthgg
			\caption{}
			\label{label_fig_tthgg}
		\end{subfigure}
		\caption{\sf \it Top: the two triangle topologies, related by $l_1\leftrightarrow l_2$,
			$\mu\leftrightarrow\nu$. Bottom: additional topologies generated by the
			non-local contact vertices, with one and two gluons attached to the non-local vertex.}
		\label{fig:feyn_diagrams}
	\end{figure}

	To capture the impact of a finite shape of the Higgs boson, the pointlike top Yukawa interaction  ($\bar t t h$) may be replaced by a momentum-dependent three-point vertex form factor,
	\begin{equation}
		V_{\ttbar h}(p',p)=-i\frac{\mt}{v}\,\Gamma_0(p',p),
		\label{eq:V0}
	\end{equation}
	where $p$ ($p'$) is the momentum of the incoming (outgoing) top quark and $q=p-p^\prime$ that of the outgoing
	Higgs boson. To recover the SM coupling at the weak scale the form factor is normalized to $\Gamma_0=1$ in the limit $p^2,p^{\prime 2}\to m_t^2$.  Such an insertion does not, in general, define a gauge-invariant
	amplitude.  It becomes evident in the top-loop contribution to $g^a(l_1)+g^b(l_2)\to h(q)$, where $l_{1,2}$ are the momenta of the on-shell external gluons and $a,b$ denote the $SU(3)_c$ color indices\footnote{Similar arguments also hold for the $h\to \gamma\gamma$ amplitude.}. 
	We schematically write the amplitude of the triangle-loop diagrams contributing to $gg \to h$, shown in Figs.~\ref{label_fig_tth_one} and \ref{label_fig_tth_two}, as
	\begin{equation}
		\mathcal A_\triangle^{ab}=\delta^{ab}\epsilon_{1\mu}\epsilon_{2\nu}\cM_\triangle^{\mu\nu},
		\label{eq:strippedAmplitude}
	\end{equation}
	where $\cM^{\mu\nu}_\triangle$ denotes the color- and polarization-stripped amplitude. Gauge invariance requires
	$l_{1\mu}\cM_\triangle^{\mu\nu}=0$, and $l_{2\nu}\cM_\triangle^{\mu\nu}=0$. The presence of the form factors, however, leaves a non-zero residual\footnote{Here $S(k)\equiv i/(\slashed k-\mt)$ denotes the top propagator, $\mathrm{tr}_D$ denotes the Dirac trace and the loop integral measure is given by 
		$\int_k\equiv\mu^{4-d}\!\int\dd^dk/(2\pi)^d$ using dimensional regularization.} 
	\begin{align} 
		\!\! l_{1\mu}\cM_{\triangle}^{\mu\nu}\propto{}
		\int_k\operatorname{tr}_{D}[S(k-l_2)\gamma^\nu S(k)]
		\Delta_{l_1}\Gamma_0(k-l_2,k),
		\label{eq:triangle-failure}
	\end{align}
	where for any vertex function $X$, we have introduced the finite-difference operator
	$\Delta_l X(p',p;\ldots)\equiv X(p',p+l;\ldots)-X(p'-l,p;\ldots)$. This explicitly displays the
	pathology of inserting a form factor arbitrarily into an otherwise local set of diagrams\footnote{In Appendix~\ref{app:1}, we demonstrate via explicit loop calculation, how the violation of the Ward identity manifests itself in the $gg\to h$ amplitude, using a monopole form factor.}.  
	%This is analogous to the violation of electromagnetic current conservation arising from off-shell hadronic form factors \cite{Ohta:1989,Haberzettl:1997jg}. 
	Apart from the SM like scenario where $\Gamma_0$ is a constant, an exception arises when $\Gamma_0(p',p)=f[(p-p')^2]=f(q^2)$, implying the two terms in
	Eq.~\eqref{eq:triangle-failure} becomes equal and cancel.
	For a generic momentum dependence, gauge invariance can instead be restored
	systematically by introducing the higher-point contact vertices $\bar t t h g$ and $\bar t t h gg$ with vertex functions,
	\begin{align}
		V_{\ttbar hg}^{\mu,a}(p^\prime, p; l)=-i g_s\frac{\mt}{v}T^a\Gamma_1^\mu(p',p;l)\,,\qquad
		V_{\ttbar hgg}^{\mu\nu,ab}(p^\prime, p; l_1,l_2)=-i g_s^2\frac{\mt}{v}\Gamma_2^{\mu\nu,ab}(p',p;l_1,l_2)\,,
		\label{eq:contactRules}
	\end{align}
	with all gluon momenta incoming. The $t\bar{t}hgg$ vertex is given by two color orderings as
	\begin{align}
		\Gamma_{2}^{\mu\nu,ab}(p^\prime,p;l_1,l_2)=T^aT^b\Gamma_{12}^{\mu\nu}(p^\prime,p;l_1,l_2)
		+T^bT^a\Gamma_{21}^{\mu\nu}(p^\prime,p;l_1,l_2),
		%\quad \Gamma_{21}^{\mu\nu}(p^\prime,p;l_1,l_2)=\Gamma_{12}^{\nu\mu}(p^\prime,p;l_2,l_1),
		\label{eq:WL2ordered}
	\end{align}
	and we define for later convenience $\Gamma^{\mu\nu}_2 \equiv \Gamma^{\mu\nu}_{12} + \Gamma^{\mu\nu}_{21}$. For $gg \to h$ amplitude at one loop, these higher-point contact interactions introduce three diagrams shown in Figs. \ref{label_fig_tthg_one}, \ref{label_fig_tthg_two} and \ref{label_fig_tthgg} in addition to the usual triangle contribution shown in Figs. \ref{label_fig_tth_one} and \ref{label_fig_tth_two}.
	
	% \begin{figure}[t]
		% \centering
		% \includegraphics[width=\linewidth]{Feynman_diags.pdf}
		% \caption{Top: the two triangle topologies, related by $p_1\leftrightarrow p_2$,
			% $\mu\leftrightarrow\nu$. Bottom: representative contact topologies generated by the
			% non-local contact vertices, with one and two gluons attached to the non-local vertex. Their
			% longitudinal parts cancel those of the triangles, while their transverse parts are model dependent.}
		% \label{fig:feyn_diagrams}
		% \end{figure}
	The total color and polarization stripped amplitude is, therefore, given by
	\begin{equation}
		\cM^{\mu\nu}=\cM^{\mu\nu}_\triangle+\cM^{\mu\nu}_{[1]}+\cM^{\mu\nu}_{[2]}+\cM^{\mu\nu}_{[12]},
		\label{eq:fiveDiagrams}
	\end{equation}
	where $\cM_{[1]}$, $\cM_{[2]}$  and $\cM_{[12]}$ corresponds to the amplitudes of Figs.~\ref{label_fig_tthg_one}, \ref{label_fig_tthg_two}, and \ref{label_fig_tthgg}, respectively. The modified Ward identity now reads as
	\begin{equation}
		l_{1\mu}\big(\cM_\triangle+\cM_{[1]}\big)^{\mu\nu}=0,\qquad
		l_{1\mu}\big(\cM_{[2]}+\cM_{[12]}\big)^{\mu\nu}=0,
		\label{eq:diagramCancellation}
	\end{equation}
	and analogously for $l_{2\nu}$ with $1\leftrightarrow2$.
	The above equations, demonstrating the restoration of gauge invariance, hold provided the contact vertices satisfy the recursive Ward--Takahashi identities (WTI)
	\begin{align}
		l_\mu\Gamma_1^\mu(p',p;l)&=\Delta_l\Gamma_0(p',p),
		\label{eq:WTI1}\\
		l_{1\mu}\Gamma_2^{\mu\nu}(p',p;l_1,l_2)&=\Delta_{l_1}\Gamma_1^\nu(p',p;l_2),
		\label{eq:WTI2a}\\
		l_{2\nu}\Gamma_2^{\mu\nu}(p',p;l_1,l_2)&=\Delta_{l_2}\Gamma_1^\mu(p',p;l_1).
		\label{eq:WTI2b}
	\end{align}
	%\repl{Upon closing the top loop, both color orderings of the two-gluon vertex yield $\Tr(T^a T^b)=\Tr(T^b T^a)=T_F\delta_{ab}$, rendering the ordering of $a,b$ immaterial. Thus,}\ab{where} the WTIs in Eqs.~\eqref{eq:WTI2a} and \eqref{eq:WTI2b} are expressed in terms of the color-stripped vertex function $\Gamma_2^{\mu\nu}$.
	
	%The double contractions of Eqs.~\eqref{eq:WTI2a} and \eqref{eq:WTI2b} give the same second finite difference of $\Gamma_0$, proving their compatibility.  Equations~\eqref{eq:WTI1}--\eqref{eq:WTI2b} then make the loop cancellation telescopic and ensure that the complete five-diagram amplitude, unlike the generic modified triangle pair, obeys Eq.~\eqref{eq:amplitudeWI}.

	% % % % % % % % % % % % % % % % % % % % % % % % % % % % % % % % % % % % % %

	Following the Ball-Chiu prescription~\cite{Ball:1980ay} we define
	\begin{align}
		\Gamma_{1,\mathrm{BC}}^\mu
		&=P_l^\mu\Delta_l^{(p)}\Gamma_0
		+P_l^{\prime\mu}\Delta_l^{(p')}\Gamma_0,
		\label{eq:BC1}\\
		\Gamma_{2,\mathrm{BC}}^{\mu\nu}
		&=P_{l_1}^\mu\Delta_{l_1}\Gamma_1^\nu
		+P_{l_2}^\nu\Delta_{l_2}\Gamma_1^\mu
		-P_{l_1}^\mu P_{l_2}^\nu\Delta_{12}\Gamma_0,
		\label{eq:BC2}
	\end{align}
	where
	\begin{align}
		\Delta_l^{(p)}X
		\equiv X(p',p+l;\ldots)-X(p',p;\ldots),
		\quad
		\Delta_l^{(p')}X
		\equiv X(p',p;\ldots)-X(p'-l,p;\ldots),
	\end{align}
	with $\Delta_l\equiv\Delta_l^{(p)}+\Delta_l^{(p')}$,
	$\Delta_{12}\equiv\Delta_{l_1}\Delta_{l_2}$, and
	$P_l^\mu=(2p^\mu+l^\mu)/(2p\cdot l)$, $P_l^{\prime\mu}
	=(2p'^\mu-l^\mu)/(2p'\cdot l)$.
	Note that, though these minimal solutions saturate the WTIs, they are not unique.  The generic vertices can be written as
	\begin{align}
		\Gamma_1^\mu=\Gamma_{1,\mathrm{BC}}^\mu+\delta\Gamma_{1}^\mu,
		\qquad
		\Gamma_2^{\mu\nu}=\Gamma_{2,L}^{\mu\nu}+\delta\Gamma_{2}^{\mu\nu},
		\label{eq:generalTransverse}
	\end{align}
	where the $\Gamma^\mu_{1,{\rm BC}}$ and $\Gamma^{\mu\nu}_{2,L}$ satisfy the corresponding recursive WTIs. In contrast, the residual $\delta\Gamma$ terms are restricted by the conditions $l_\mu\delta\Gamma_{1}^\mu=0$, and $l_{1\mu}\delta\Gamma_{2}^{\mu\nu}=l_{2\nu}\delta\Gamma_{2}^{\mu\nu}=0$. For on-shell gauge bosons, $\delta\Gamma$ terms can contain a polarization-null part, proportional to the external gluon momenta, and a non-null transverse part. Schematically,
	\begin{align}
		\delta\Gamma_{1}^\mu=l^\mu \Gamma_{1,{\rm null}}
		+\Gamma_{1,\perp}^\mu,\quad
		\delta\Gamma_{2}^{\mu\nu}=l_1^\mu l_2^\nu \Gamma_{2,{\rm null}}
		+\Gamma_{2,\perp}^{\mu\nu}.
		\label{eq:nullPhysical_1}
	\end{align}
	The WTIs cannot determine either $\Gamma_{i, {\rm null}}$ or $\Gamma_{i,\perp}$, however, the null terms, being proportional to the gluon momenta do not contribute to any physical process with on-shell gluons. In contrast, the non-null transverse components may leave a physical contribution to processes like $gg \to h$, since  
	$\epsilon_\mu\Gamma_{1,\perp}^\mu\ne0$, and $
	\epsilon_{1\mu}\epsilon_{2\nu}\Gamma_{2,\perp}^{\mu\nu}\ne0$.%The WTIs cannot determine either $\Gamma_{i, {\rm null}}$ or $\Gamma_{i,\perp}$, however, the latter may leave a physical contribution to the $gg \to h$ amplitude.  
	The Ball-Chiu prescription amounts to setting these unconstrained terms to zero, which can be determined within specific ultraviolet theories. Similar Ball-Chiu type $\bar t t h$ vertices also contribute to the $h\to \gamma\gamma$ process when the gluons are replaced by the photons.
	
	% % % % % % % % % % % % % % % % % % % % % % % % % % % % % % % % % % % % % %
	\section{Gauge invariant non-local action}
	\label{non-local}
	% % % % % % % % % % % % % % % % % % % % % % % % % % % % % % % % % % % % % %
	
	A transparent way to obtain the complete tower of gauge-related
	vertices is to formulate the momentum-dependent Yukawa interaction directly
	as a gauge-invariant non-local action, by joining the colored fields
	by Wilson lines. A minimal QCD-gauge-invariant, hermitian and translation invariant non-local action in the position space is given by\footnote{Electromagnetic gauge invariance formally requires
		an additional $U(1)_{\rm EM}$ Wilson line connecting the
		spatially separated top fields. Since we are focusing on the gluon-fusion process $gg\to h$, for the time being we retain
		only its zeroth-order contribution, setting
		$\mathcal W_{\rm EM}(x,y)=1$. Electromagnetic gauge
		interactions can be incorporated systematically by
		expanding the corresponding Wilson line.}
	
	\begin{align}
		\mathcal{S}_{\rm NL}
		={}&
		-\frac{m_t}{v}
		\int d^4x\,d^4y\,d^4z\;
		\bar t(x)\,
		\mathcal{W}_3(x,y)\mathcal{K}(x-z,y-z)\,
		t(y)\,
		h(z),
		\label{eq:nonlocal-action}
	\end{align}
	where $\mathcal{K}(x-z,y-z)$ is a Lorentz-scalar non-local kernel and
	\begin{align}
		\mathcal{W}_3(x,y)
		=
		\mathcal{P}
		\exp\left[
		ig_s
		\int_{\mathcal{C}}
		d\xi^\mu\,
		G_\mu^a(\xi)T^a
		\right]
		\label{eq:qcd-wilson-line}
	\end{align}
	is a path ordered QCD Wilson line along a contour $\mathcal{C}$ connecting $y$ to $x$. %The Wilson line therefore provides the gauge completion of the spatially separated quark bilinear. 
	Under an $SU(3)_c$ gauge transformation $U(x)$, the top quark field transforms as $t\to Ut$, while the Wilson line transforms as $\mathcal W_3(x,y)\to U(x)\mathcal W_3(x,y)U^\dagger(y)$, thereby rendering the 
	$\mathcal{S}_{\rm NL}$ gauge invariant for any contour. In terms of $u\equiv x-z$ and $w\equiv y-z$, the form
	factor $\Gamma_{0}(p^\prime,p)$ is obtained through the Fourier transform of the action as
	\begin{equation}
		\Gamma_0(p',p)=\int\dd^4u\,\dd^4w\;e^{ip'\cdot u-ip\cdot w}\,\mathcal K(u,w).
		\label{eq:fourier}
	\end{equation}
	The local limit $\mathcal K=\delta^{(4)}(u)\delta^{(4)}(w)$ gives $\Gamma_0=1$, and hermiticity of
	$\mathcal{S}_{\rm NL}$ implies $\Gamma_0(p',p)^*=\Gamma_0(p,p')$ for real momenta away from singularities. The choice of contour encodes additional information about the underlying non-local dynamics, beyond that contained in $\Gamma_0$ alone.
	Translation and Lorentz invariance require the contour to be built
	from $u$ and $w$ alone, so it lies in the plane of the triangle $(x,y,z)$. Such a contour can be parametrized as: $\xi(\tau)=z+a(\tau)\,u+b(\tau)\,w$, where $\tau=0$ ($\xi =y$) and $\tau=1$ ($\xi=x$) correspond to the endpoints of the contour. The path-dependent functions $a(\tau)$ and $b(\tau)$ specify the shape of the contour with the boundary conditions $(a,b)\big|_{\tau=0}=(0,1)$, and $(a,b)\big|_{\tau=1}=(1,0)$.
	Expanding $\mathcal{W}_3$ in $g_s$ one finds
	\begin{align}
		\Gamma_1^\mu(p',p;l)&=\int_0^1\!\dd\tau\;\mathcal D^\mu(\tau)\,\Gamma_0\big(p'-a\,l,\,p+b\,l\big),
		\qquad \mathcal D^\mu(\tau)\equiv\frac{da}{d\tau}\,\partial_1^\mu-\frac{db}{d\tau}\,\partial_2^\mu,
		\label{eq:WL1}\\
		\Gamma_{12}^{\mu\nu}(p',p;l_1,l_2)&=\int_0^1\!\dd\tau_1\!\int_0^{\tau_1}\!\dd\tau_2\;
		\mathcal D^\mu(\tau_1)\,\mathcal D^\nu(\tau_2)\,
		\Gamma_0\big(p'-a(\tau_1)l_1-a(\tau_2)l_2,\,p+b(\tau_1)l_1+b(\tau_2)l_2\big),
		\label{eq:XX12}
	\end{align}
	where $\partial^\mu_1$ ($\partial^\mu_2$) denotes derivative with respect to the first (second) argument of $\Gamma_0$. It is straightforward to verify that the above expressions reproduce the recursive WTIs upon contraction with either gluon momenta. Analogously, imposing electromagnetic gauge invariance generates the corresponding higher-point contact vertices, which satisfy their respective recursive WTIs. Note that changing the integration path while keeping its endpoints fixed leaves the WTIs unchanged, but may modify the transverse structure of the vertices.
	
	While the Ball-Chiu construction saturates the WTIs, we consider two representative contours to illustrate the possible transverse structures that remain unconstrained by these identities. The direct contour $\mathcal C_D:y\to x$ has $(a,b)=(\tau,1-\tau)$,
	and is natural when the interaction is viewed as a bilocal quark operator. The broken contour
	$\mathcal C_B:y\to z\to x$ has $(a,b)=(0,1-2\tau)$ for $\tau\le\tfrac12$ and $(2\tau-1,0)$ for
	$\tau\ge\tfrac12$, and is natural when the interaction arises from two non-local quark legs meeting
	at the Higgs insertion point. Equation~\eqref{eq:WL1} then gives
	\begin{align}
		\Gamma_{1,D}^\mu&=\int_0^1\!\dd\tau\,\big(\partial_1^\mu+\partial_2^\mu\big)
		\Gamma_0\big(p'-\tau l,\,p+(1-\tau)l\big),
		\label{eq:WL1direct}\\
		\Gamma_{1,B}^\mu&=\int_0^1\!\dd\tau\,\Big[\partial_1^\mu\Gamma_0\big(p'-\tau l,\,p\big)
		+\partial_2^\mu\Gamma_0\big(p',\,p+(1-\tau)l\big)\Big].
		\label{eq:WL1broken}
	\end{align}
	For the two-gluon vertex, using Eq.~\eqref{eq:XX12} with the direct and broken paths we find
	\begin{align}
		\Gamma_{D,12}^{\mu\nu}={}&\int_0^1\!\dd\tau_1\!\int_0^{\tau_1}\!\dd\tau_2\;
		\mathcal (\partial_1^\mu+\partial_2^\mu)(\partial_1^\mu+\partial_2^\mu)\,
		\Gamma_0\big(p'-a(\tau_1)l_1-a(\tau_2)l_2,\,p+b(\tau_1)l_1+b(\tau_2)l_2\big),
	\end{align}
	and
	\begin{align}
		\Gamma_{B,12}^{\mu\nu}={}&\int_0^1\!\dd\tau_1\!\int_0^{\tau_1}\!\dd\tau_2\Big[
		\partial_1^\mu\partial_1^\nu\Gamma_0\big(p'-\tau_1l_1-\tau_2l_2,\,p\big)
		+\partial_2^\mu\partial_2^\nu\Gamma_0\big(p',\,p+(1-\tau_1)l_1+(1-\tau_2)l_2\big)\Big]
		\nonumber\\
		&+\int_0^1\!\dd\tau_1\!\int_0^1\!\dd\tau_2\;
		\partial_1^\mu\partial_2^\nu\Gamma_0\big(p'-\tau_1l_1,\,p+(1-\tau_2)l_2\big).
		\label{eq:WL2broken}
	\end{align}
	In Eq.~\eqref{eq:WL2broken}, the first two terms correspond to both gluons on the same segment, while in the
	last term one gluon lies on each segment.
	
	The difference between the direct and broken contours can be
	expressed in terms of mixed momentum derivatives of the form
	factor $\partial_1^\mu\partial_2^\nu\Gamma_0$. 
	Additionally, the difference for the one-gluon vertex is a null term, which does not contribute to physical observables with external gluons. Therefore, a sufficient condition for path independence till order $\mathcal O(g_s^2)$ is given by
	\begin{equation}
		\partial_1^\mu\partial_2^\nu\Gamma_0(p',p)=0.
		\label{eq:pathindependence}
	\end{equation}
	For example, an additively separable form factor
	$\Gamma_0(p',p)=f(p')+g(p)+c$ yields identical vertices for both the contour choices. 
	%For generic nonseparable form factors, however, the contours may differ through contributions not uniquely fixed by the WTIs, while preserving gauge invariance. 

	% % % % % % % % % % % % % % % % % % % % % % % % % % % % % % % % % % % % % %

	%--------------------------------------------------------------------
	\subsection{Structure of $\Gamma_0$ from lowest resonance dominance}
	\label{sec:illustrativeFF}

	Gauge invariance determines how a given $\Gamma_0$ must be accompanied by a tower of contact vertices, but not
	$\Gamma_0$ itself, which encodes the dynamics of the strong sector. The form factor $\Gamma_0(p',p)$ can depend on
	the Lorentz invariant  momenta $p'^2$, $p^2$, and $q^2$, with particular
	momentum structures motivated by different realizations of strong
	dynamics. 
	
	A particularly interesting category is given by 
	$\Gamma_0=f[(\alpha p'+\beta p)^2]$, as encountered in
	covariant constituent-quark and non-local chiral quark
	models~\cite{GomezDumm:2006vz,Branz:2009cd}.
	The corresponding non-local kernel restricts the three spacetime
	points to a collinear configurations, and the Wilson-line
	completion becomes independent of the contour choice. A special case is $\Gamma_0=f(q^2)$ with $q^2=(p'-p)^2$, where the entire tower of gluonic
	contact vertices vanishes. Consequently, the triangle diagrams
	alone satisfy the WTIs, and the $gg\to h$ amplitude is simply
	rescaled by $f(m_h^2)$.
	A different class of form factor arises when
	$\Gamma_0=f(p'^2)g(p^2)$, which is motivated by non-local NJL and
	instanton-inspired quark models~\cite{Bowler:1994ir,Plant:1997jr,GomezDumm:2006vz}.
	Its non-vanishing mixed momentum derivatives generally lead to
	contour-dependent transverse contributions. In contrast, an
	additively separable form
	$\Gamma_0=f(p'^2)+g(p^2)$ satisfies
	$\partial_1^\mu\partial_2^\nu\Gamma_0=0$, yielding
	contour-independent contact vertices. 
	
	For the rest of the analysis, we select a symmetric, additively separable form factor.
	%\paragraph{Benchmark: lowest-resonance dominance.}
	Motivated by the resonance structure of strongly coupled
	theories, we consider the
	momentum dependence associated with each top-quark leg
	is dominated by the lightest fermionic resonance corresponding
	to a monopole form factor. Imposing symmetry
	under $p\leftrightarrow p'$ and requiring the SM normalization for on-shell top quarks, we adopt
	\begin{equation}
		\Gamma_0(p',p)
		= \frac{m_*^2-\mt^2}{2}\left[\frac{1}{\mathcal D_*(p)}+\frac{1}{\mathcal D_*(p^\prime)}\right],
		\label{eq:monopole}
	\end{equation}
	where $\mathcal{D}_*(p)=m_*^2-p^2$, and $m_*$ characterizes the mass scale of the lightest composite fermionic excitation. This
	parametrization is motivated by the familiar
	single-pole dominance of hadronic form factors~\cite{Masjuan:1210.0760}, partial compositeness framework~\cite{Agashe:2004rs,
		Barnard:2013zea} and the resonance
	saturation of correlation functions in large-$N$ \cite{tHooft:1973alw,tHooft:1974pnl,Witten:1979kh}
	strongly coupled theories~\cite{Barnard:2013zea,
		Banerjee:2105.01093}.
	
	For momentum transfers well below the resonance
	scale, the form factor admits the expansion
	\begin{equation}
		\Gamma_0(p',p)
		=1+\frac{p'^2+p^2-2\mt^2}{2m_*^2}
		+\mathcal O(m_*^{-4}),
		\label{eq:ff-eft-expansion}
	\end{equation}
	recovering the expected local derivative expansion
	as $m_*\to\infty$. However, as the momenta approach the resonance
	scale, the derivative expansion ceases to converge,
	necessitating the retention of the full momentum
	dependence of the form factor.
	
	Additive separability of the form factor ensures that the Wilson-line completion is independent of the choice of contour. The one-gluon vertex is given by the Ball-Chiu term $\Gamma^\mu_{1, {\rm BC}}$ plus a polarization-null contribution $\Gamma_{1,{\rm null}}$ (see Eq~\eqref{eq:generalTransverse}) as
	\begin{align}
		\Gamma_{1}^\mu
		={}&\frac{m_*^2-\mt^2}{2}
		\left[
		\frac{(2p+l)^\mu}
		{\mathcal D_*(p)\mathcal D_*(p+l)}
		+\frac{(2p'-l)^\mu}
		{\mathcal D_*(p')\mathcal D_*(p'-l)}
		\right] +l^\mu \Gamma_{1,{\rm null}},
		\label{eq:BC1-monopole}
	\end{align}
	while the two-gluon vertex takes the form
	\begin{align}
		\Gamma_2^{\mu\nu}
		={}(m_*^2-\mt^2)
		\int_0^1\!\dd\sigma_1
		\int_0^1\!\dd\sigma_2
		\Big[
		H^{\mu\nu}(p'-\sigma_1l_1-\sigma_2l_2)
		+H^{\mu\nu}(p+\sigma_1l_1+\sigma_2l_2)
		\Big],
		\label{eq:G12-monopole}
	\end{align}
	where
	\begin{equation}
		H^{\mu\nu}(P)
		=\frac{g^{\mu\nu}}{(P^2-m_*^2)^2}
		-\frac{4P^\mu P^\nu}{(P^2-m_*^2)^3}.
		\label{eq:H-def}
	\end{equation}
	Unlike the one-gluon case, the two-gluon Wilson
	vertex contains a nontrivial transverse contribution
	relative to the minimal Ball-Chiu prescription.
	For $gg\to h$, the resulting difference between
	the Ball-Chiu prescription and the Wilson-line action based gauge completion enters through the
	two-gluon contact diagram in Fig.~\ref{label_fig_tthgg}.

	\section{Loop induced Higgs processes}
	\label{pheno}
	% % % % % % % % % % % % % % % % % % % % % % % % % % % % % % % % % % % % % %
	
	For the $gg\to h$ process the amplitude may be written as
	\begin{equation}
		\cM_g^{\mu\nu}=i \frac{g_s^2T_F}{8\pi^2 v} A_{t}(s,m_*)T^{\mu\nu},\quad
		T^{\mu\nu}=\frac{s}{2}g^{\mu\nu}-l_2^\mu l_1^\nu.
		\label{eq:tensor}
	\end{equation}
	%Here $\mathcal C_V$ contains the channel-dependent coupling and color or charge factors.
	For the non-local scenario, $A_{t}(s,m_*)$ is given as
	\begin{equation}
		A_t(s,\ms)
		=
		A^{\rm BC}_t(s,\ms)+A^T_t(s,\ms),
	\end{equation}
	where we separate the contribution from the model dependent transverse part of the $t\bar{t}hgg$ vertex into $A^T_t(s,\ms)$, and denote the rest of the contributions including the Ball-Chiu vertices satisfying the WTIs into $A^{\rm BC}_t(s,\ms)$. Using the conventional Passarino--Veltman (PV) function $C_{abc}\equiv C_0(0,0,s;a^2,b^2,c^2)$, we obtain $A^{\rm BC}_t(s,\ms)$ and $A^T_t(s,\ms)$ as
	\begin{align}
		A^{\rm BC}_t(s,\ms)
		&=
		-\frac{4m_t^2}{s}
		\left[
		4\ms^2C_{m_*m_*m_*}
		-
		sC_{m_tm_tm_*}
		+
		(s-4\mt^2)C_{m_tm_tm_t}
		\right],\label{eq:amp_1}\\
		A^T_t(s)
		& =
		\frac{8m_t^2(\ms^2-m_t^2)}{s^2}
		\int_0^1\dif x\,
		\Li\left[
		\frac{x(1-x)s}
		{x\ms^2+(1-x)\mt^2-\ii0}
		\right]
		-
		\frac{8m_t^2}{s}
		\left[
		\Delta B_0-4\Delta C_{00}
		\right].
		\label{eq:amp_2}
	\end{align}
	The definitions of the PV integrals, including the
	ultraviolet-finite combinations $\Delta B_0$ and
	$\Delta C_{00}$, are collected in
	Appendix~\ref{app:PVconventions}.
	%The loop integrals were evaluated using \texttt{Package-X}~\cite{Patel:2015tea,Patel:2016fam} interfaced with \texttt{CollierLink}~\cite{Denner:2016kdg}.}

The total amplitude is finite since the divergent part of all $B_0$ and $C_{00}$ functions cancel in the sum. The decoupling of the non-local effects are evident at large $m_*$, as $A^{\rm BC}_t\to A^{\rm SM}_t$ and $A^T_t\to 0$ when $m_*\to \infty$. At large $m_*$, the pure transverse term scales as $A^T_t\sim 1/m_*^2$, while the Ball-Chiu contribution becomes logarithmically enhanced\footnote{The present framework illustrates a
	departure from the conventional Higgs low-energy theorem
	(LET), which relates the effective Higgs--gluon coupling
	to the Higgs dependence of the heavy-fermion mass matrix
	~\cite{Shifman:1979eb,Kniehl:1995tn,Low:2010mr,
		Gillioz:2012se}. In minimal composite-Higgs models,
	this relation can render gluon fusion largely insensitive
	to the top-partner spectrum
	~\cite{Falkowski:2007hz,Azatov:2011qy,Montull:2013mla,Chen:2014xwa}.
	Here, the non-local top--Higgs interaction is modified
	while the SM top propagator is retained, placing the
	effective description outside the conventional LET
	assumptions.} as $A^{\rm BC}_t\sim 1/m_*^2\ln\left(m_t^2/m_*^2\right)$, numerically, $|A^T_t|/|A^{\rm BC}_t|\sim 10^{-3}$ ($10^{-4}$) for $m_*=5\, (10)$ TeV. The log enhancement can be traced back to the momentum dependent Ball-Chiu vertex. To elucidate this we rewrite the low momentum expansion of the vertex function given in Eq.~\eqref{eq:ff-eft-expansion} as a position space effective local Lagrangian
\begin{align}
	\mathcal{L}_{\rm eff} \sim -\frac{m_t}{v}\bar{t} t h -\frac{m_t}{2vm_*^2} \bar{t}\left[(i D)^2-m_t^2\right]t h +{\rm h.c.}\supset \frac{g_s m_t}{2vm_*^2}\bar t \sigma^{\mu\nu}t^a G^a_{\mu\nu} t h.
\end{align}
This chromomagnetic dipole term imply a log enhancement in the amplitude due to the underlying momentum dependent form factors in non-local theories. This may be contrasted with the parametric suppression of the transverse contribution arising from the $\bar{t}t h gg$ vertex. %which is specific to the minimally Wilson-line dressed action prescribed in Eq.~\eqref{eq:nonlocal-action}. 
%However, one could envisage a gauge invariant non-local action dressed with a similar chromomagnetic gluon field strength operator ($\bar t(x) \mathcal{W}_3(x,r)\sigma^{\mu\nu}G_{\mu\nu}(r) \mathcal{W}_3(r,y)t(y) h(z)$), which expectedly lead to a logarithmic enhancement.

Neglecting the sub-percent bottom loop, the signal strength for gluon-fusion process is therefore
\begin{equation}
	\mu_{gg}=\frac{\sigma(gg\to h)}{\sigma(gg\to h)_{\mathrm{SM}}}
	=\frac{|A_{t}(\mh^2,m_*)|^2}{|A_{t}^{\mathrm{SM}}(\mh^2)|^2}.
	\label{eq:mugg}
\end{equation}
where $A^{\rm SM}_t(m_h^2)$ is given in \cite{Djouadi:2005gi}. For diphoton amplitude, including the interference with the $W$ loops, which is left at its SM value \cite{Djouadi:2005gi}, the signal strength is given by
\begin{equation}
	\mu_{\gamma\gamma}=\frac{\Gamma(h\to \gamma\gamma)}{\Gamma(h\to \gamma\gamma)_{\mathrm{SM}}}=
	\frac{|N_cQ_t^2 A_{t}(\mh^2,m_*)+A_W^{\mathrm{SM}}(\mh^2)|^2}
	{|N_cQ_t^2 A_{t}^{\mathrm{SM}}(\mh^2)+A_W^{\mathrm{SM}}(\mh^2)|^2}.
	\label{eq:mugamma}
\end{equation}
The monopole form factor suppresses the top-loop driven amplitude causing $\mu_{gg}<1$, while it in turn weakens the destructive
$W$--top interference and enhances $h\to\gamma\gamma$ decay width. In Appendix~\ref{app:1}, we compare $\mu_{gg}$ obtained from the fully gauge-invariant formulation with that from a na\"ive form-factor insertion in the $\bar t t h$ vertex alone in which the Ward identity is enforced by hand.
\begin{figure*}[t!]
	\centering
	\includegraphics[width=0.48\textwidth]{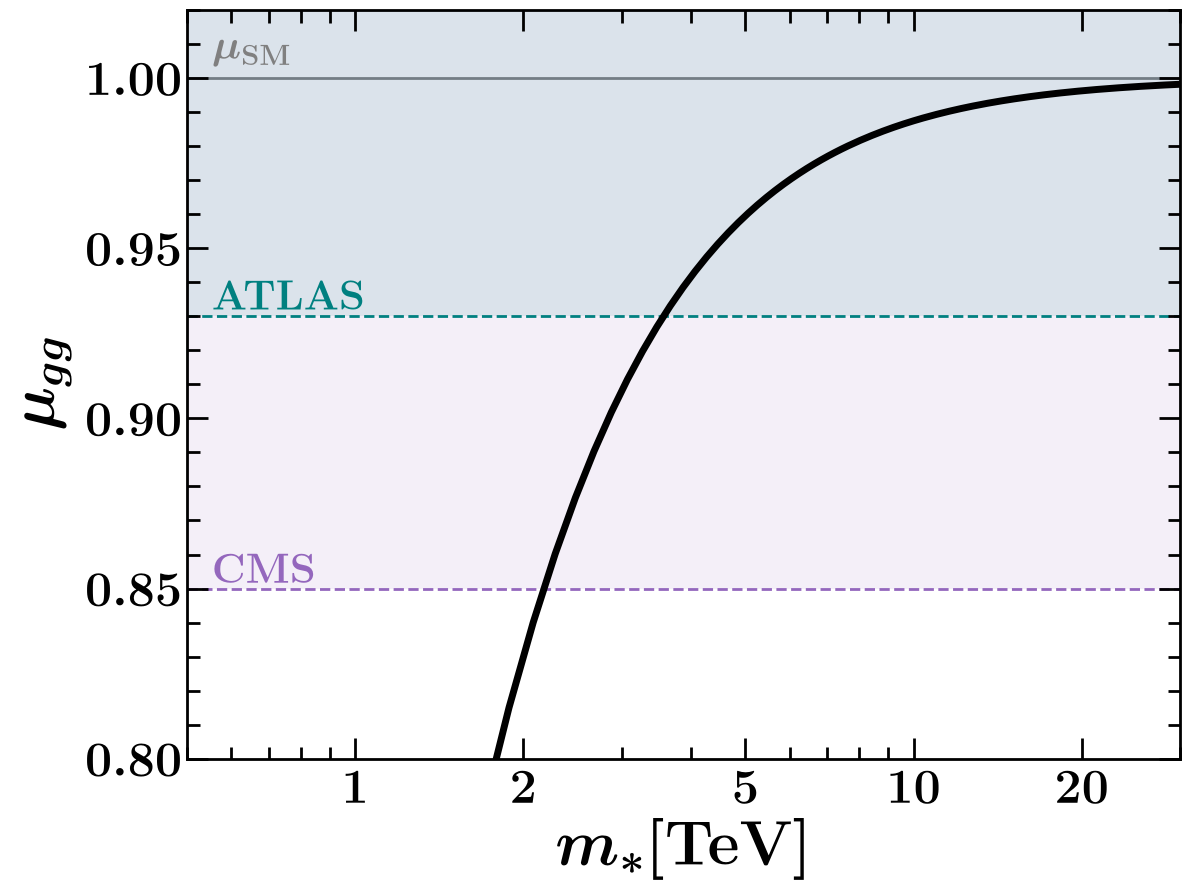}
	\includegraphics[width=0.48\textwidth]{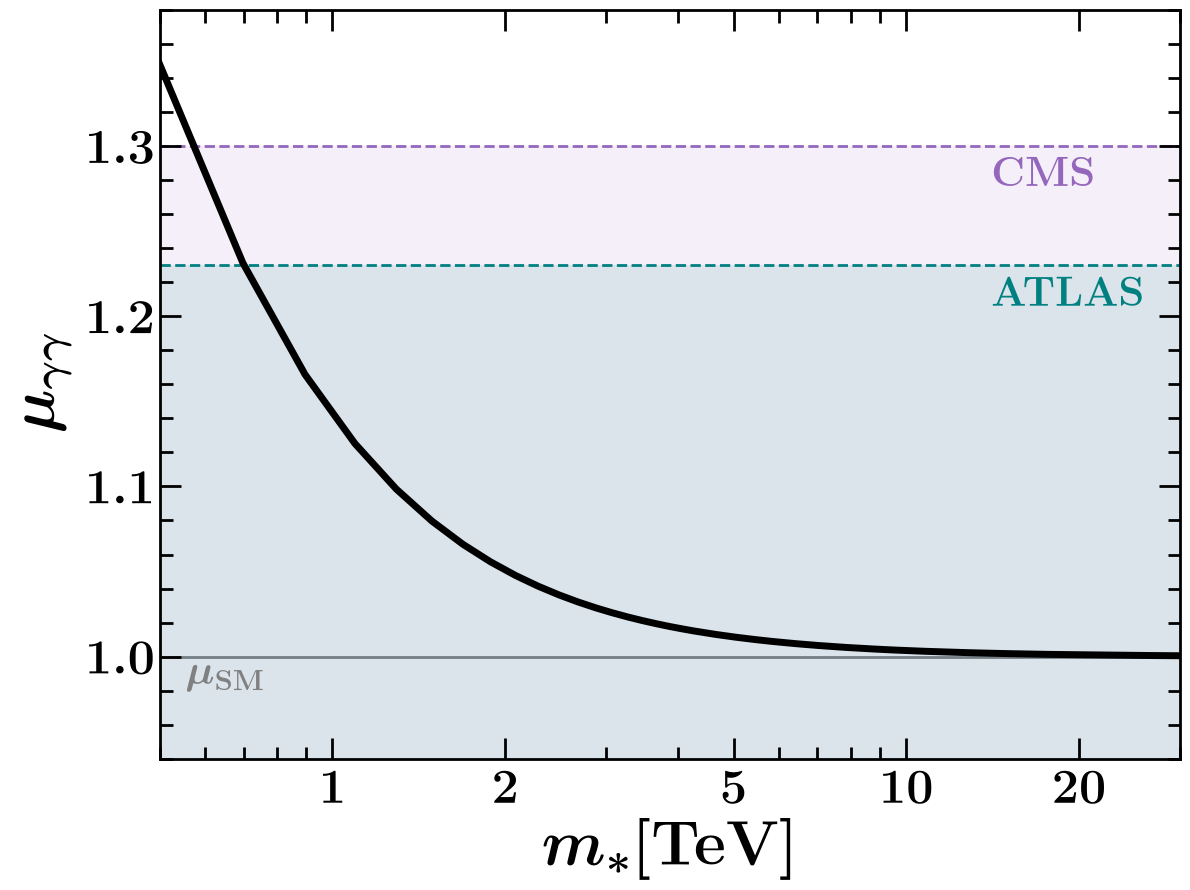}
	\caption{\sf\it Gluon-fusion (left) and diphoton (right) signal
		strengths as functions of the resonance mass $m_*$,
		obtained using the monopole top--Higgs form factor.
		The black solid curves represent the theoretical
		predictions, while the shaded areas
		indicate the regions allowed by current experimental constraints from the LHC Run 2 data.}
	\label{fig_mu_hgg}
\end{figure*}    
\begin{table}[!]
	\centering
	\small
	\setlength{\tabcolsep}{9pt}
	\renewcommand{\arraystretch}{1.25}
	\begin{tabular}{lcc|cc}
		\toprule
		& \multicolumn{2}{c|}{$gg\to h$}
		& \multicolumn{2}{c}{$h\to\gamma\gamma$} \\
		\cmidrule(lr){2-3}
		\cmidrule(lr){4-5}
		Experiment
		& $\delta\kappa_g$ (\%)
		& $m_*$ [TeV]
		& $\delta\kappa_\gamma$ (\%)
		& $m_*$ [TeV] \\
		\midrule
		HL-LHC ~\cite{ATLASCMS:2025HLLHC}
		& 2.4 & 2.97
		& 1.8 & 1.58 \\
		
		ILC250~\cite{deBlas:2019rxi}
		& 2.3 & 3.05
		& 6.7 & 0.60 \\
		
		CEPC~\cite{deBlas:2019rxi}
		& 1.5 & 3.97
		& 3.7 & 0.95 \\
		
		FCC-ee/eh/hh~\cite{deBlas:2019rxi}
		& 0.49 & 7.73
		& 0.29 & 4.99 \\
		\bottomrule
	\end{tabular}
	\caption{
		\sf\it Projected future-collider sensitivities to the resonance scale $m_*$. 
		The quantities $\delta\kappa_g$ and $\delta\kappa_\gamma$ denote the 
		projected uncertainties in the effective Higgs couplings within the 
		$\kappa$ framework. These are translated into signal-strength 
		uncertainties and compared with the predictions of the momentum-dependent 
		top--Higgs form factor, assuming SM central values for the signal 
		strengths.
	}
	\label{tab:future_higgs_sensitivity}
\end{table}

The CMS and ATLAS collaborations have reported the following measurement of the signal strengths using Run 2 data with integrated luminosity of 140 fb$^{-1}$: $(\mu^{\rm CMS}_{gg}, \mu^{\rm CMS}_{\gamma\gamma})=(0.99^{+0.08}_{-0.07}, 1.12\pm 0.09)$, and $(\mu^{\rm ATLAS}_{gg},\mu^{\rm ATLAS}_{\gamma\gamma})=(1.05^{+0.07}_{-0.06},1.05\pm 0.09)$~\cite{CMS:2026Combined,ATLAS:2025Combined}.
The variations of $\mu_{gg}$ and $\mu_{\gamma\gamma}$ on the resonance scale $m_*$
are shown in Fig.~\ref{fig_mu_hgg}, together with the current
ATLAS and CMS measurements.
Given these measurements and the specific pattern of deviations in the minimal model, the strongest constraint arises from
the ATLAS measurement of $\mu_{gg}$, excluding $m_*\lesssim
3.6$~TeV at the $2\sigma$ level. The projected sensitivities
of future facilities, including the HL-LHC, ILC, CEPC, and FCC,
are summarized in Table~\ref{tab:future_higgs_sensitivity}
\cite{ATLASCMS:2025HLLHC,deBlas:2019rxi}.

This may be compared with the direct LHC searches for pair-produced top partners which currently
probe masses of order $1.5$~TeV, depending on their decay modes
and branching fractions; see Ref.~\cite{Banerjee:2024VLQ}
for a review. Although the precise relation between $m_*$
and the physical resonance spectrum is model dependent,
these results demonstrate that
loop-induced Higgs processes can provide competitive
constraints on the underlying strong dynamics.
The gluon-fusion channel generally offers stronger sensitivity
because the top-quark loop provides the dominant SM contribution,
whereas in $h\to\gamma\gamma$ the effect of the modified top loop
is moderated by the dominant $W$-boson contribution, with which
it interferes destructively.

\begin{figure*}[t!]
	\centering
	\includegraphics[width=0.48\textwidth]
	{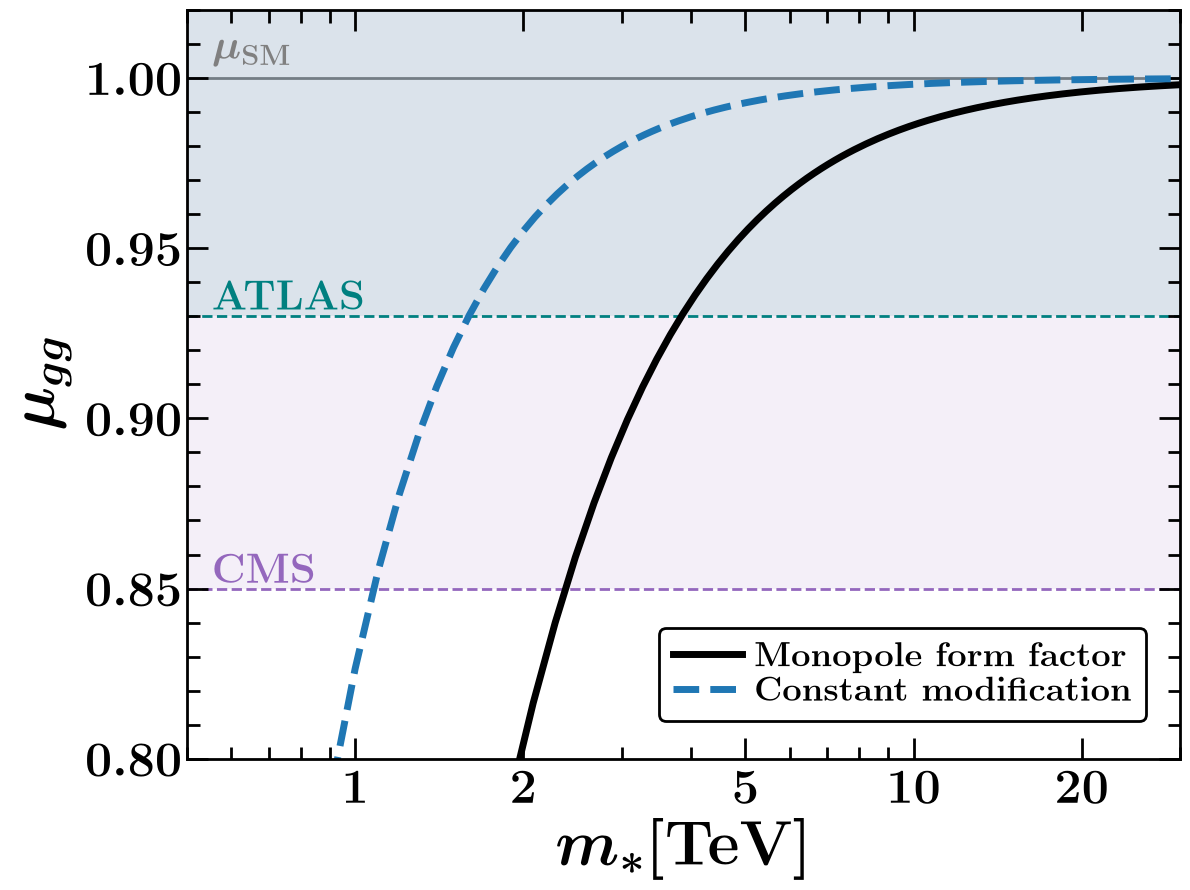}
	\includegraphics[width=0.48\textwidth]
	{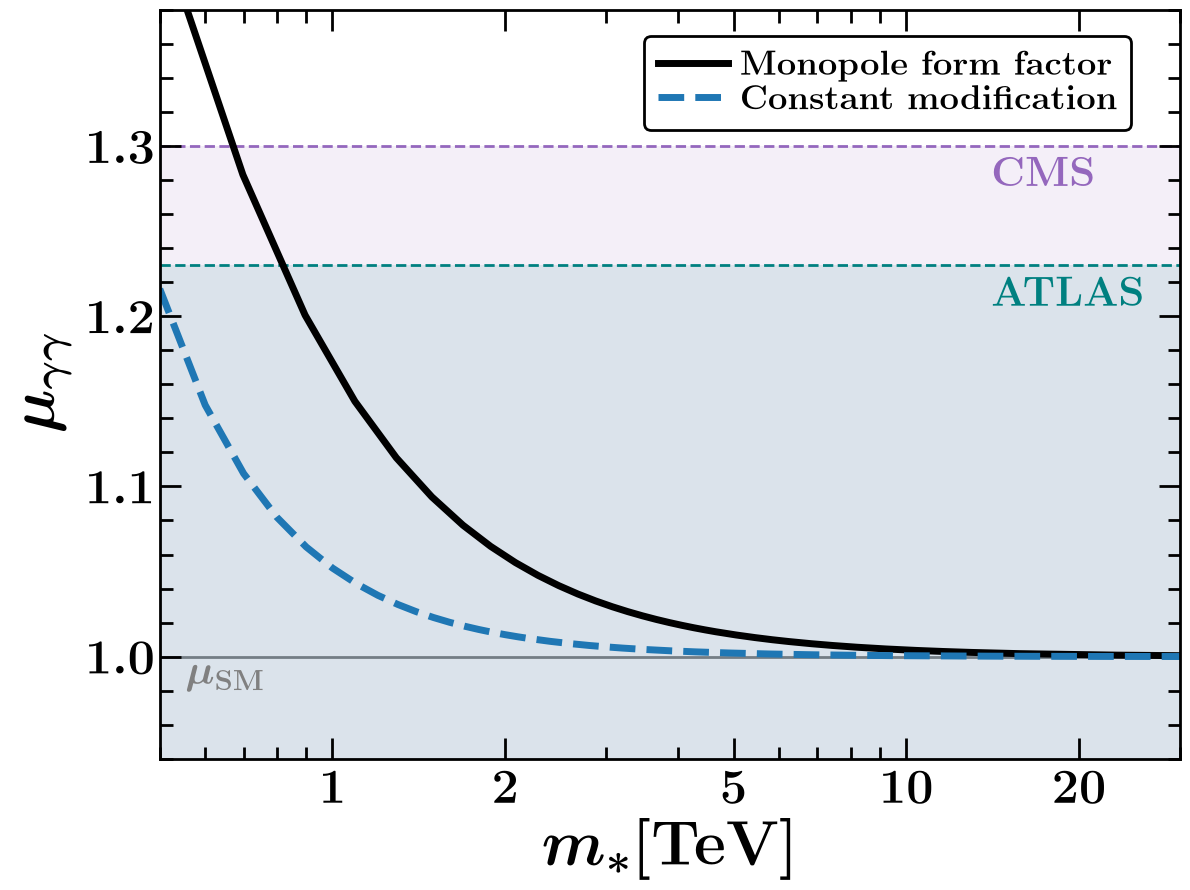}
	\caption{\sf\it Comparison of the gluon-fusion (left) and
		diphoton (right) signal strengths as functions of the
		fermionic resonance mass $m_*$ (with $g_*=1$), for the illustrative
		composite-Higgs benchmark scenario.
		The black solid curves represent the full gauge-invariant
		prediction obtained from the momentum-dependent monopole
		form factor, while the
		blue dashed curves correspond to a momentum-independent
		top-Yukawa rescaling by $\kappa_t$.}
	\label{fig_mu_hgg_compare}
\end{figure*}

To illustrate the effects of the pseudo-Nambu--Goldstone boson (pNGB)
nature of the Higgs in composite Higgs scenarios, we extend
our benchmark form factor by incorporating the universal
rescaling of the top Yukawa coupling characteristic of a
representative minimal composite Higgs ~\cite{Montull:2013mla,Banerjee:2017wmg,Banerjee:2020tqc,Bautista:2020mxw}:
\begin{align}
	\Gamma_0(p^\prime,p)
	=\kappa_t\frac{m_*^2}{2}\left[\frac{1}{\mathcal D_*(p)}+\frac{1}{\mathcal D_*(p^\prime)}\right],
	\qquad
	\kappa_t=\frac{1-2\xi}{\sqrt{1-\xi}},
	\qquad
	\xi\equiv\frac{v^2}{f^2}.
	\label{eq:composite-monopole}
\end{align}
Here, $f$ denotes the pNGB Higgs decay
constant, which sets the scale of global symmetry breaking. The characteristic mass of the fermionic
resonances is related to this scale through $m_*=g_*f$,
where $g_*$ represents the effective strong-sector coupling.

Figure~\ref{fig_mu_hgg_compare} compares the gluon-fusion
and diphoton signal strengths obtained from the complete
momentum-dependent form factor with those predicted by a
constant top-Yukawa rescaling by $\kappa_t$, as commonly adopted
in composite Higgs phenomenology. For the illustrative
choice $g_*=1$, the constant-rescaling approximation yields
considerably weaker sensitivity to the resonance scale
$m_*$ than the full non-local interaction. This highlights
the importance of retaining the momentum dependence of the
top--Higgs vertex, which probes virtualities within the loop
and captures effects that cannot be reproduced by a
momentum-independent coupling modifier.

% % % % % % % % % % % % % % % % % % % % % % % % % % % % % % % % % % % % % %
\section{Conclusion}
\label{sec:conclusion}
% % % % % % % % % % % % % % % % % % % % % % % % % % % % % % % % % % % % % %

In this work, we present an approach  toward consistently incorporating a momentum-dependent top-Higgs form factor into the loop-induced amplitudes $gg \to h$ and $h \to \gamma\gamma$. When the $t\bar{t}h$ vertex depends on the virtualities of the top quarks, a simple form factor insertion into the standard triangle diagrams violates the Ward-Takahashi identities. To restore gauge invariance,  we formulate a non-local Yukawa action dressed with a Wilson line and match its expansion to a minimal Ball-Chiu-type longitudinal vertex.  For an additively separable symmetric monopole form factor motivated by lowest-resonance dominance, the Wilson-line completion becomes contour-independent and allows us to evaluate the complete one-loop amplitudes in closed analytic form.  We find that the  Ball-Chiu contribution yields a finite, logarithmically enhanced deviation from the Standard Model scaling as $(m_t^2/m_*^2)\ln(m_*^2/m_t^2)$. Within this  framework  the parametrically suppressed  transverse components of the multi-gluon contact vertices remain unconstrained  and in general depend on the choice of the Wilson-line contour.

Since  the loop integral samples virtualities up to the new physics scale and beyond,  this framework is more sensitive to the compositeness scale than a momentum-independent $\kappa_t$ rescaling. The resulting deviations are correlated, suppressing the gluon-fusion production rate ($\mu_{gg} < 1$) while simultaneously enhancing the diphoton decay width ($\mu_{\gamma\gamma} > 1$) via reduced destructive interference with the $W$-boson loop. Using the current LHC signal-strength measurements, we obtain a $2\sigma$ exclusion of the new physics scale below $3.6~\text{TeV}$, while future precision projections across the HL-LHC, ILC, CEPC, and FCC are expected to probe scales up to $8~\text{TeV}$. While the present study focuses on loop-induced inclusive processes with an on-shell Higgs boson, the non-local $t\bar{t}h$ vertex and its accompanying $t\bar{t}hg$ and $t\bar{t}hgg$ contact interactions will also enter directly at tree level in channels with higher kinematic thresholds, which will be taken up in a subsequent study.
\section*{Acknowledgments}
% % % % % % % % % % % % % % % % % % % % % % % % % % % % % % % % % % % % % %

A.B. was supported by the Department of Atomic Energy, Government of India, during the initial
stage of this project. R.P. is supported by the National Natural Science Foundation of China (12425506 and 12375101). T. S. R. acknowledges the Department of Science and Technology, Government of India, under the ANRF Grant Agreement No. MTR/2023/000469 (MATRICS) for financial assistance. We thank Agni Keyoor Purani and Samanwaya Patra for discussions. A.B. thanks International Centre for Theoretical Sciences (ICTS) for hospitality during the program - School for Advanced Topics in Particle Physics (SATPP): Selected Topics in Effective Field Theories (code: ICTS/SATPP2024/04), and especially Matthias Neubert for his lectures on the method of regions, which rekindled our interest in this project. The authors acknowledge assistance from OpenAI GPT and Anthropic's Claude with parts of the numerical analysis, presentation and proofreading of the manuscript. 
%The authors reviewed all outputs and take full responsibility for the content of the manuscript and its results.

% % % % % % % % % % % % % % % % % % % % % % % % % % % % % % % % % % % % % %
\appendix
% % % % % % % % % % % % % % % % % % % % % % % % % % % % % % % % % % % % % %

% % % % % % % % % % % % % % % % % % % % % % % % % % % % % % % % % % % % % %
\section{Triangle-only amplitude and Ward identity violation}
\label{app:1}
% % % % % % % % % % % % % % % % % % % % % % % % % % % % % % % % % % % % % %

To illustrate the necessity of a gauge-invariant treatment of
momentum-dependent interactions, we examine the consequences
of inserting a non-local top--Higgs form factor directly into
the conventional triangle diagrams for $gg\to h$, without
including the associated gluonic contact interactions.
This calculation serves two purposes: it explicitly
demonstrates how the violation of the Ward identities
manifests itself in the loop amplitude and quantifies
the error that may arise when gauge invariance is
subsequently imposed by hand, rather than derived
from a consistent underlying action.

For definiteness, we consider the symmetric monopole
form factor introduced in the main text,
\begin{equation}
	\Gzero(p',p)
	=\frac{m_*^2-\mt^2}{2}\left[\frac{1}{\mathcal D_*(p)}+\frac{1}{\mathcal D_*(p^\prime)}\right].
	\label{eq:app-ff}
\end{equation}
Retaining the tensor structures relevant for physical
on-shell gluon polarizations, the triangle contribution
can be decomposed as
\begin{equation}
	\cM_{\triangle}^{\mu\nu}
	=
	A_\triangle(s,\ms)\,g^{\mu\nu}
	+
	B_\triangle(s,\ms)\,l_2^\mu l_1^\nu,
	\label{eq:app-decomp}
\end{equation}
where $l_1^2=l_2^2=0$ and
$s=(l_1+l_2)^2=2l_1\cdot l_2$.
The coefficients obtained from the form-factor insertion
are
\begin{align}
	A_\triangle
	&=
	\frac{i g_s^2(\mt^2-\ms^2)\mt}{16\pi^2 v}
	\int_0^1 dx
	\int_0^{1-x}dy
	\int_0^1 dz\,
	\frac{
		z^2\left[
		2\mt^2+s\left(-1+2xz[1+(y-1)z]\right)
		\right]
	}{
		\mathcal D^2(x,y,z)
	},
	\label{eq:app-A}
	\\[2mm]
	B_\triangle
	&=
	\frac{i g_s^2(\mt^2-\ms^2)\mt}{8\pi^2 v}
	\int_0^1 dx
	\int_0^{1-x}dy
	\int_0^1 dz\,
	\frac{
		z^2\left[
		1-4xz(1+(y-1)z)
		\right]
	}{
		\mathcal D^2(x,y,z)
	},
	\label{eq:app-B}
\end{align}
with
\begin{equation}
	\mathcal D(x,y,z)
	=
	\ms^2(z-1)-\mt^2 z
	+sxz\left[1+(y-1)z\right].
	\label{eq:app-D}
\end{equation}

Gauge invariance requires the amplitude to be transverse
with respect to both external gluon momenta.
For the tensor decomposition in
Eq.~\eqref{eq:app-decomp}, contraction with the
external momenta yields
\begin{align}
	l_{1\mu}\cM_\triangle^{\mu\nu}
	=R_\triangle(s,\ms)\,l_1^\nu,\quad 
	l_{2\nu}\cM_\triangle^{\mu\nu}
	=R_\triangle(s,\ms)\,l_2^\mu,\quad
	{\rm where}\quad
	R_\triangle(s,\ms)
	\equiv
	A_\triangle(s,\ms)
	+\frac{s}{2}B_\triangle(s,\ms).
	\label{eq:app-residual-def}
\end{align}
Substituting Eqs.~\eqref{eq:app-A}
and \eqref{eq:app-B}, we obtain
\begin{equation}
	R_\triangle(s,\ms)
	=
	\frac{i g_s^2(\mt^2-\ms^2)\mt}{8\pi^2 v}
	\int_0^1 dx
	\int_0^{1-x}dy
	\int_0^1 dz\,
	\frac{
		z^2\left[
		\mt^2-sxz\left(1+(y-1)z\right)
		\right]
	}{
		\mathcal D^2(x,y,z)
	}
	\neq 0.
	\label{eq:app-residual}
\end{equation}
The nonvanishing residual explicitly exhibits the
failure of the triangle-only contribution to satisfy
the Ward identities.

One might nevertheless assume that the form factor
originates from some unspecified underlying strong
dynamics whose additional contributions restore gauge
invariance. A na\"ive approach is to enforce this condition
algebraically while retaining the calculated
coefficient $A_\triangle$ and replacing
$B_\triangle$ by $-2A_\triangle/s$.
The resulting amplitude is
\begin{equation}
	\cM_{\triangle}^{\mu\nu}
	=
	\frac{2}{s}A_\triangle(s,\ms)\,
	T^{\mu\nu},
	\qquad
	T^{\mu\nu}
	=
	\frac{s}{2}g^{\mu\nu}
	-l_2^\mu l_1^\nu.
	\label{eq:app-naive-amplitude}
\end{equation}
We choose to retain $A_\triangle$ rather than
$B_\triangle$ as the independent coefficient,
with the prescription intended to reproduce the
SM amplitude in the decoupling limit
$\ms\to\infty$.

Although Eq.~\eqref{eq:app-naive-amplitude} is
manifestly transverse, this procedure is not
equivalent to calculating a gauge-invariant
amplitude obtained from the Wilson-line formulation. 

\iffalse
Enforcing transversality
by hand cannot account for the contributions
of the additional interactions required by
the underlying dynamics. In particular, such
interactions may modify the coefficient of the
gauge-invariant tensor itself, an effect entirely
missed by the na\"ive prescription.
\fi

\begin{figure*}[t!]
	\centering
	\includegraphics[width=0.5\textwidth]
	{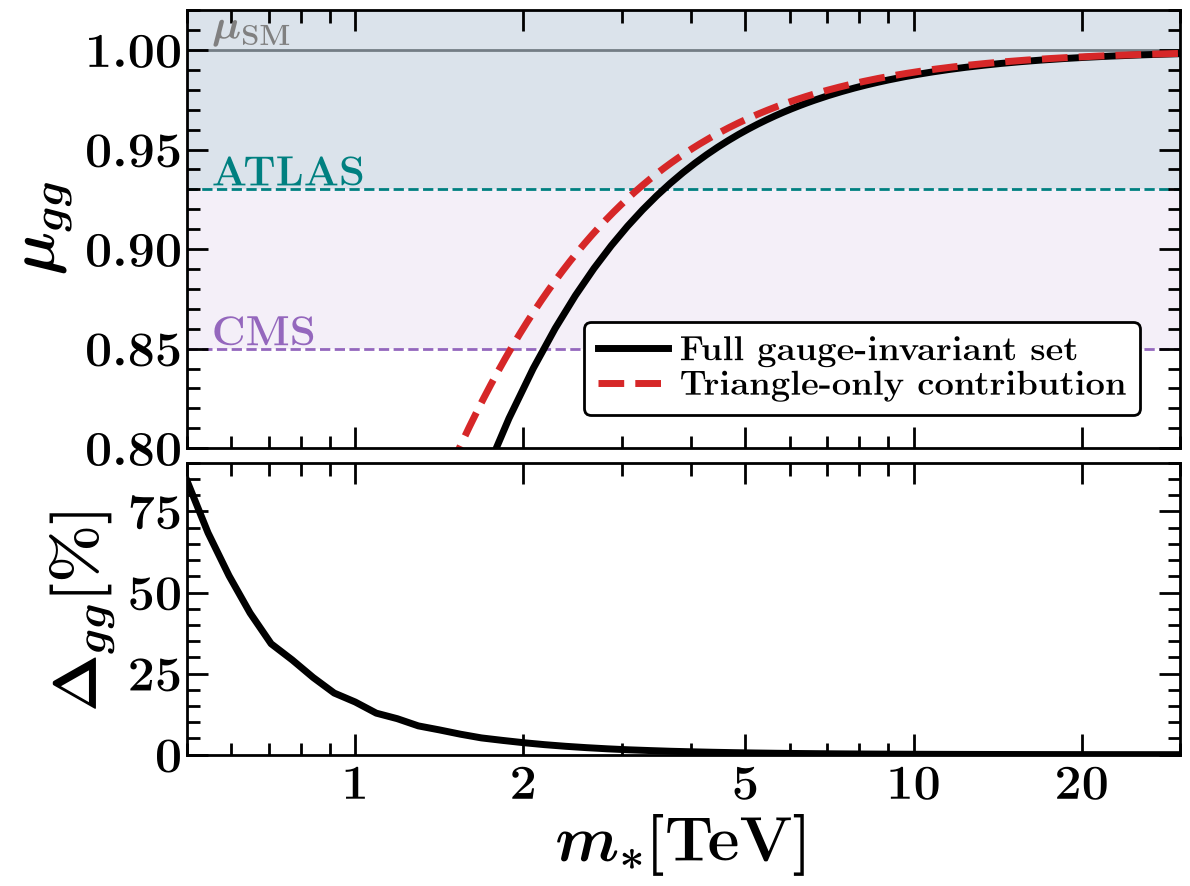}
	\caption{\sf\it
		Comparison of the gluon-fusion signal strength
		$\mu_{gg}$ as a function of the fermionic resonance
		mass $m_*$. The black solid curve
		represents the complete gauge-invariant prediction
		obtained from the Wilson-line formulation, while
		the red dashed curve corresponds to the
		triangle-only calculation with the Ward identity
		enforced algebraically. Despite satisfying the same
		transversality condition, the two prescriptions
		differ by approximately $5\%$ at $m_*=2~\mathrm{TeV}$,
		illustrating the impact of the additional interactions
		required by gauge invariance.
	}
	\label{fig:comparison}
\end{figure*}

%Our Wilson-line formulation resolves this ambiguity by constructing the gluonic contact interactions directly from a gauge-invariant non-local action. These interactions contribute to the loop amplitude and ensure that the Ward identities are satisfied by the complete set of diagrams, without imposing additional constraints on the calculated tensor coefficients.

%To assess the phenomenological impact of neglecting these contributions, 
We compare the gluon-fusion
signal strength obtained from the
Ward-enforced triangle amplitude with the
fully gauge-invariant prediction. As shown in
Fig.~\ref{fig:comparison}, the two calculations
yield different results even though both
amplitudes are transverse. In particular, the
relative difference in the signal strength,
\begin{equation}
	\Delta_{gg}(\ms)
	\equiv
	\left|
	\frac{\mu_{gg}^{\triangle}(\ms)}
	{\mu_{gg}^{\mathrm{full}}(\ms)}
	-1
	\right|,
	\label{eq:app-relative-error}
\end{equation}
is approximately $5\%$ at $\ms=2~\mathrm{TeV}$
for our benchmark.

This comparison demonstrates that restoring the
Ward identities algebraically does not guarantee
the correct prediction for a loop-induced observable.
The discrepancy arises because the triangle-only calculation
fails to capture the complete contributions associated
with the momentum-dependent interaction. 

\iffalse
Although its
precise magnitude depends on the chosen form factor
and gauge-invariant completion, the example highlights
the importance of calculating the full set of
interactions dictated by gauge invariance, rather
than relying on an ad hoc restoration of the
Ward identities.
\fi 

% % % % % % % % % % % % % % % % % % % % % % % % % % % % % % % % % % % % % %
\section{Conventions for the Passarino--Veltman functions}
\label{app:PVconventions}
% % % % % % % % % % % % % % % % % % % % % % % % % % % % % % % % % % % % % %

We collect here the conventions for the one-loop scalar and tensor
integrals entering Eqs.~\eqref{eq:amp_1} and \eqref{eq:amp_2}. We work in dimensional
regularization with $d=4-2\epsilon$ with the normalization $\mathcal N_d\equiv (2\pi\mu)^{4-d}/i\pi^2$, where $\mu$ denotes the dimensional-regularization scale. 
%These definitions follow the conventional normalization employed by \textsc{Collier}~\cite{Denner:2016kdg}.
For $p^2=s$, the scalar two-point function is
\begin{align}
	B_0(s;a^2,b^2)
	&=\mathcal N_d\int d^d k\,
	\frac1{[k^2-a^2][(k+p)^2-b^2]}.
	\label{eq:PV_B0_def}
\end{align}
For two massless external gauge bosons with momenta $p_1$, $p_2$, and $s=(p_1+p_2)^2$, the scalar three-point function is defined as 
\begin{equation}
	C_{abc}\equiv C_0(0,0,s;a^2,b^2,c^2)
	=\mathcal N_d\int d^d k\,\frac1{D_aD_bD_c}.
	\label{eq:PV_C0_def}
\end{equation}
where we define the three propagator denominators as
\begin{align}
	D_a=k^2-a^2,\quad
	D_b=(k+p_1)^2-b^2,\quad
	D_c=(k+p_1+p_2)^2-c^2.
	\label{eq:PV_Denominators}
\end{align}
The rank-two three-point integral is decomposed as
\begin{align}
	C^{\mu\nu}(0,0,s;a^2,b^2,c^2)
	\equiv \mathcal N_d\int d^d k\,\frac{k^\mu k^\nu}{D_aD_bD_c}
	=g^{\mu\nu}C_{00}+p_1^\mu p_1^\nu C_{11}
	+p_2^\mu p_2^\nu C_{22}
	+\bigl(p_1^\mu p_2^\nu+p_2^\mu p_1^\nu\bigr)C_{12}.
	\label{eq:PV_C00_def}
\end{align}
The combinations entering the transverse contributions (see Eq.~\eqref{eq:amp_2}) to the $gg \to h$ amplitude are
\begin{align}
	\Delta B_0\equiv B_0(s;m_t^2,m_*^2)-B_0(s;m_*^2,m_*^2),
	\quad
	\Delta C_{00}\equiv
	C_{00}(0,0,s;m_t^2,m_*^2,m_*^2)
	-C_{00}(0,0,s;m_*^2,m_*^2,m_*^2).
	\label{eq:PV_finite_differences}
\end{align}
Each difference is ultraviolet finite, since the poles in
$B_0$ and $C_{00}$ do not depend on their internal masses.
Consequently, $\Delta B_0-4\Delta C_{00}$ is independent of
the ultraviolet subtraction convention and of $\mu$. We performed the analytic reduction of the one-loop integrals
using \texttt{Package-X}~\cite{Patel:2015tea,Patel:2016fam}.
The numerical evaluation of the PV scalar and tensor coefficients
was carried out using \texttt{CollierLink}, the \texttt{Mathematica}
interface to the \texttt{Collier} library~\cite{Denner:2016kdg}.

\bibliography{nonlocal_top_higgs}
\bibliographystyle{JHEP}
% % % % % % % % % % % % % % % % % % % % % % % % % % % % % % % % % % % % % %
\end{document}